# Forces and symmetry breaking of a living meso-swimmer

R. A. Lara[‡], N. Sharadhi[‡], A. A. L. Huttunen, L. Ansas, E. J. G. Rislakki, G. M. Bessa, and M. Backholm*
Department of Applied Physics, Aalto University, Espoo, Finland
*Email: matilda.backholm@aalto.fi; [‡]These authors contributed equally.

Swimming is ubiquitous in nature and crucial for the survival of a wide range of organisms. The physics of swimming at the viscosity-dominated microscale and inertia-dominated macroscale is well studied. However, in between lies a complicated mesoscale with swimmers affected by non-linear and time-dependent fluid mechanics. The intricate motility strategies, combined with complex and periodically changing body shapes add extra challenges for accurate meso-swimming modelling. Here, we have further developed the micropipette force sensor to directly probe the swimming forces of the meso-organism *Artemia*. Through deep neural network-based image analysis, we show how *Artemia* achieves an increased propulsive force by increasing its level of time-reversal symmetry breaking. We present a universal force-based scaling law for a wide range of micro- to meso-organisms with different body shapes, swimming strategies, and level of inertia at the mesoscale. These results capture fundamental aspects of biological meso-swimming dynamics and provide guidance for future biomimicking meso-robot designs.

## Introduction

Motility, or the ability of an organism to move, plays a crucial role in the survival, adaptation, and evolution of various organisms (*1*). In liquid, swimming is the main form of locomotion where organisms need to displace the surrounding fluid through periodic modulations of their bodies to propel forward. Numerous different swimming strategies exist in a wide range of species and length scales. Understanding the physics of biological swimming provides guidance for future bio-inspired robots with vast biomedical applications (*2*). The type of swimming motion commonly used at the macroscale does not necessarily render net propulsion at the microscale (*3*). From a physical perspective, a successful motility strategy is determined by the interplay between inertial ($F_\mathrm{i}$) and viscous ($F_\eta$) forces, as expressed by the dimensionless Reynolds number $\mathrm{Re} = F_\mathrm{i}/F_\eta \sim \rho U L/\eta$ (**Fig. 1A**). Here, $L$ is the length scale of the swimmer, $U$ its swimming velocity, and $\rho$ and $\eta$ is the fluid density and viscosity. For small organisms, such as bacteria, algae, and nematodes, $\mathrm{Re} \ll 1$ and viscous forces dominate. Conversely, large organisms, such as humans, fish, and dolphins, move in an inertial regime at $\mathrm{Re} \gg 1$. Due to the linearity and time-reversibility of Stokes' equation at low Re, micro-swimmers need to break time-reversal symmetry through non-reciprocal motion to achieve net propulsion, as defined by the Scallop theorem (*3*).

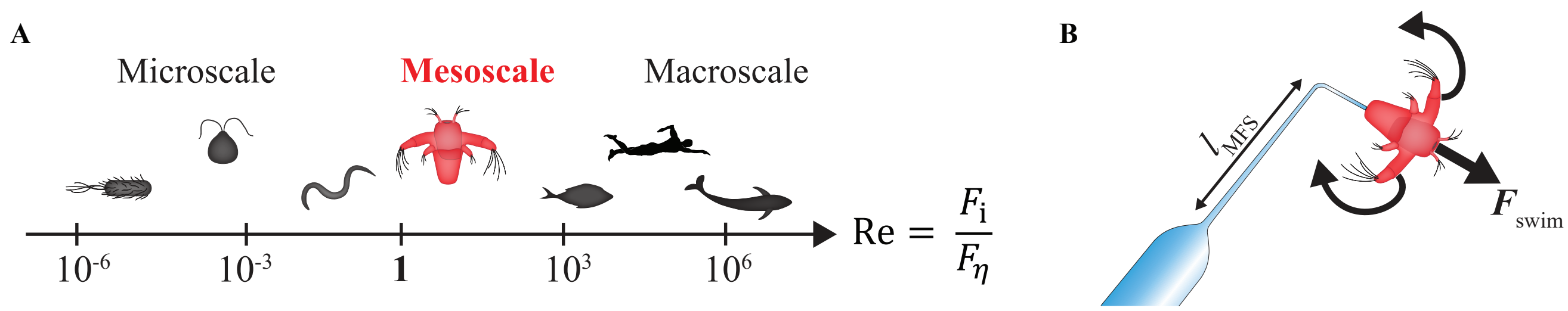

**Fig. 1 | Swimming dynamics of living organisms. A)** The Reynolds number of organisms with different length scales. The physics behind swimming is well understood at the micro- and macroscale. In the intermediate mesoscale, *Artemia* (red) has evolved to achieve propulsion in a complicated fluid mechanics regime governed by non-linearities and time-dependencies. **B)** We have performed direct swimming force ($F_\mathrm{swim}$) measurements using a micropipette forcer sensor to probe the butterfly-swimming dynamics of *Artemia*. The cantilever (length $l_\mathrm{MFS}$, schematic not to scale) is custom-made and calibrated in our lab to fit the requirements of the experiments.

Between the micro- and macroscale lies an intermediate mesoscale (**Fig. 1A**) where both viscous and inertial forces are important (*4*). Animals at $\mathrm{Re} \approx 1 - 1000$ and body sizes of $L \approx 0.5$ mm to 50 cm fall within this regime. The mesoscale thus hosts many different organisms (*5–12*), such as small larvae, shrimp, and jellyfish, as well as rudimentary artificial swimmers, such as magneto-capillary dumbbells (*13*) and two-sphere meso-swimmers (*14*). These meso-swimmers are affected by various levels of inertial, nonlinear, and time-dependent effects (*4*, *13*), all of which are expected to create complex propulsion strategies and dynamics (*15*). Theoretical work has shown that mesoscale motility often is counterintuitive in nature and sensitive to the parameter space of finite inertia (*16*). An important example is that the Scallop theorem breaks down at the mesoscale where the addition of inertia allows for reciprocal swimming motions (*13*, *16–21*). Theory suggests that inertia starts inducing changes in the swimming gaits of physical model-swimmers (e.g., a flapping plate) at $\mathrm{Re} \gtrsim 5 - 20$ (*21–23*), whereas the speed of freely swimming organisms transition to the mesoscale (i.e., from the Stokes to the laminar regime) at $\mathrm{Re} > 10^2$ (*24*). Despite ongoing research, it remains unclear how the swimming dynamics of real living organisms are influenced by the interaction between viscous and inertial forces at the mesoscale and whether they exploit the breakdown of the Scallop theorem in their locomotion.

At low and high Re, simplifications can be made to the Navier-Stokes equations to model the locomotion of simply shaped objects (*21*, *25*). However, such assumptions of purely viscous or inviscid fluid dynamics cannot be used at the mesoscale where both inertial and viscous forces are of similar magnitude (*21*). Furthermore, analytical complications arise with organisms whose shapes cannot be approximated as spheres or cylinders (*26*), which is the case for many living meso-organism that have developed into complex, time-varying shapes. Various experimental and analytical approaches have been made to describe the swimming speed and frequencies across different length scales (*24*, *27–30*). Furthermore, custom-made computational fluid dynamics software has become a powerful tool for modelling swimming dynamics (*31–36*), where the geometry and muscle activation of the organism, as well as the fluid-structure interaction during its swimming are taken into account. The currently available literature for intermediate Re swimmers is primarily focused on nematodes (*31*), small fishes (*32*), and larvae (*33*), while the most studied swimming strategies are undulatory (*31–33*) and metachronal (*34*, *35*). The main limitations of these models relate to the complex geometries and fluid-structure interactions of non-spherical or non-cylindrical organisms. To address this, direct force measurements are proposed as a synergistic technique that can provide quantitative data on organisms with complex shapes and motions (*26*). The integration of modelling and experimental approaches enables a more comprehensive understanding of locomotion, providing quantitative insights into their dynamics, efficiency, strength, and physiology. However, force measurements on small swimming organisms is challenging and rely on advanced experimental approaches (*37*). Moreover, the drag experienced by a non-swimming organism passively moving through liquid cannot be accurately compared to that of an actively swimming one, as the thrust-generating appendages (e.g., tail, flagella, or antennae) induce complex and time-dependent fluid flow on the non-thrust-producing body (*26*). Several force-based techniques, including the optical tweezer (*38–42*), micropipette force sensor (*43–47*), semiconductor force sensor (*5*), and different spring-based probes (*6*, *10*, *12*), have been developed to measure tiny swimming forces (~$10^{-12}$ to $10^{-6}$ N) of, for example, individual bacteria (*38*, *39*), algae (*40*, *41*, *47*), spermatozoa (*42*), nematodes (*43–46*, *48*), ascidian larvae (*10*), corixidae (*12*), and copepods (*5*, *6*). Currently, it is still unclear exactly how the propulsion dynamics differ between organisms of different lengths, morphologies, swimming strategies, and Re.

In this study, we have performed direct force measurements on a living meso-swimmer to study the effects of increasing inertia at intermediate Re (**Fig. 1B**). We have further developed the micropipette force sensor (MFS) technique (*46*) to directly probe the fast (~10 Hz) swimming forces of *Artemia sp*., a widely utilized live feed for fish and aquatic invertebrates as well as a model meso-swimmer (*49*, *50*). *Artemia* hatches at a fairly low $\mathrm{Re} \approx 2$ where viscous forces often can be assumed to still dominate (*51*), but transitions into an intermediate $\mathrm{Re} \approx 10 - 40$ as it grows, accompanied by a change in motility strategy from butterfly swimming to metachronal gliding (*49*, *52*). To date, the mesoscale size of *Artemia* has rendered direct probing of its swimming forces difficult. The previous tour-de-force measurements by Williams were conducted exclusively on a large-scale, Re-matched physical model, constructed from brass and tungsten wires (*50*). Although useful for basic force studies, this is not expected to capture the intricate kinematics and dynamics of a real *Artemia*. Using the MFS technique, we have managed to perform a detailed and direct analysis of the mesoscale dynamics of this meso-swimmer at different life stages. By using deep neural network-based image analysis of the *Artemia* kinematics, we quantify the level of non-reciprocity of the swimmers. We identify a correlation between the propulsive force and the level of time-reversal symmetry breaking of the swimming motion, rendering a quantitative demonstration of the importance of non-reciprocity in micro- to mesoscale swimming. Finally, we gather the current swimming force measurement data from the literature and show a universal scaling law for swimming dynamics spanning the full range of micro- to mesoscale organisms, all with different morphologies and swimming strategies.

## Results and Discussion

### ***Developing MFS for swimming force experiments on Artemia***

In the MFS technique, the spring-like deflection ($x$, **Fig. 2A–B**, **Supplementary Movie 1**),) of a long, thin, and hollow glass micropipette is used to measure forces with a sub-nN resolution (*46*). In our work, we have manufactured L-shaped cantilevers following a standard protocol (*46*) (see *Methods*). Tethered swimming is a widely used method to measure swimming forces (*5*, *6*, *10*, *12*, *53*), although the tethers may influence the swimming dynamics (*54*). The acceleration reaction, where the mass of water entrained by the swimming organ is maximally accelerated backwards, has shown to not significantly contribute to the thrust force at the start of the power stroke of swimmers at Re $< 100$ (*12*, *50*). The force generated by the added mass is thus not an important source of propulsion at the lower end of the intermediate Re relevant to *Artemia* swimming. Furthermore, we minimise any hydrodynamic interactions between the swimmer and the cantilever by making the bent tip of the MFS cantilever (**Fig. 2A**) sufficiently long (ca. 1 mm). We also carefully compare the tethered swimming kinematics and its surrounding flow fields with that of free swimmers to evaluate the effect of the force sensor.

In most previous work where MFS has been used to probe forces in soft and/or living mesoscale systems (e.g., Refs. (*43–45*, *55–57*)), the measurements were deemed quasi-static and the cantilever drag and inertia was neglected. In such case, the force was determined as $F = kx$, where $k$ is the elastic spring constant of the cantilever. In our work, the cantilever motion cannot be regarded as quasi-static since *Artemia* moves with high swimming frequencies and powerful swimming strokes, rendering large and fast pipette deflections. Frostad *et al*. (*58*) and Böddeker *et al*. (*47*) have developed different approaches to allow for more dynamic measurements with cantilever-based techniques, and we have followed the former in our work. The swimming force of *Artemia* in our MFS measurements can thus be determined by considering the elastic, viscous, and inertial forces of the pipette: $F_{\text{swim}} = kx + b\dot{x} + m_{\text{eff}}\ddot{x}$, where $b$ and $m_{\text{eff}}$ are the viscous drag coefficient and effective mass of the MFS (*58*). The three coefficients were determined through calibration (see *Methods*). A quasi-static water drop calibration technique (*46*, *55*) was used to measure $k$ (**Supplementary Note 2: Supp. Fig. S1**), which decreases strongly with increasing cantilever length (**Fig. 2C**), enabling the fabrication of MFSs with varying stiffness.

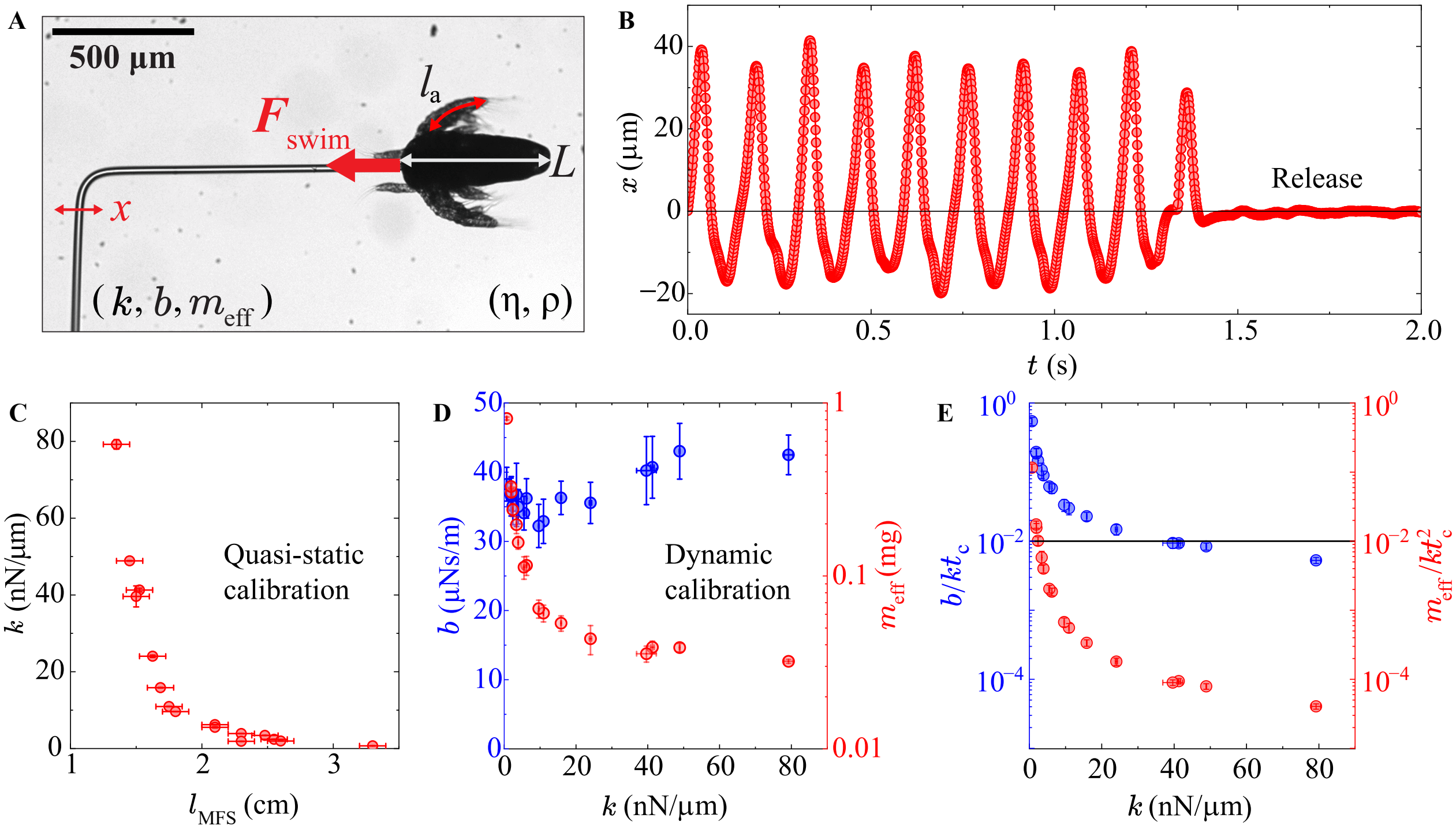

**Fig. 2 | Micropipette force sensor for swimming force measurements on *Artemia*. A)** Optical microscopy image from an MFS swimming force experiment on *Artemia* (body length $L = 520 \pm 30\ \mu$m, antenna length $l_a$) caught by the head through suction. The *Artemia* swims in a brine medium (viscosity $\eta$, density $\rho$). A high-speed camera captured the pipette deflection ($x$) as well as the motion of the *Artemia* body during several swimming cycles. To determine $F_{\text{swim}}$ from $x$, the cantilever needs to be calibrated for the spring constant $k$, damping coefficient $b$, and effective mass $m_{\text{eff}}$. **B)** Example of MFS deflection data as a function of time. At ca. 1.4 s, the organism is released to determine the equilibrium, zero-force MFS position. **C)** MFS cantilever spring constant (obtained through the quasi-static water drop calibration method) as a function of its length. **D)** The MFS damping coefficient (blue data, left *y*-axis) and effective mass (red data, right *y*-axis) obtained through the dynamic calibration method as a function of spring constant. **E)** The relative contribution of cantilever drag vs. elasticity (blue data, left *y*-axis) and inertia vs. elasticity (red data, right *y*-axis) as a function of spring constant. A critical threshold of 1% (solid line) is used to determine for which $k$ the effects of drag and inertia must be considered. The error bars in C–E are the standard deviations of repeated measurements on the same pipette.

To determine $b$ and $m_{\text{eff}}$, the MFS was calibrated with a dynamic approach where the fluid-immersed cantilever was deflected and released to oscillate back to its equilibrium position (**Supplementary Movie 2, Supp. Fig. S2A**, see *Methods*). The pipette deflection was tracked as a function of time with a high-speed camera and fit with a damped harmonic oscillator model $x(t) = Ae^{-\xi\omega_o t}\sin\left(\sqrt{1-\xi^2}\omega_o t + \varphi\right)$ (*58*, *59*) (**Supp. Fig. S2B**). From the fitting parameters (amplitude $A$, phase $\varphi$, damping ratio $\xi = b/2\sqrt{m_{\text{eff}}k} = b\omega_0/2k$, undamped oscillation frequency $\omega_0$) of this model, the drag coefficient $b = 2k\xi/\omega_0$ and effective mass $m_{\text{eff}} = k/\omega_0^2$ could be determined for MFSs of different lengths (**Fig. 2D**). The effective mass decreases as a function of increasing $k$ (decreasing $l_{\text{MFS}}$), whereas the damping coefficient remains constant within experimental error. We verify that the measured magnitudes of $b$ and $m_{\text{eff}}$ are reasonable through additional calculations (see **Supplementary Note 1** for more details). Based on the characteristic timescale of the *Artemia* swimming cycle ($t_c \approx 0.1$ s, **Fig. 2B**), the contribution of MFS drag compared to elasticity ($\sim b/kt_c$) and inertia compared to elasticity ($\sim m_{\text{eff}}/kt_c^2$) was calculated (*60*) (**Fig. 2E**). For our cantilevers, drag effects are considered for $0.7 < k < 25$ nN µm$^{-1}$, while the contribution of inertia is deemed negligible. To avoid resonance effects (**Supp. Fig. S2C–D**), MFSs in the range of $5 < k < 8$ nN µm$^{-1}$ were not used for the swimming force experiments.

***Tethered and free-swimming kinematics***

To perform swimming force measurements, several *Artemia* with body lengths of $L = 420 - 1500$ µm were captured at either their heads or the abdomen ("tail") with an L-shaped MFS using gentle suction. The organism was aligned to swim orthogonally to the cantilever (**Fig. 1B** and **2A**, see *Methods*). In this work we focused on *Artemia* swimming with a "butterfly" motion where the secondary antennae move back and forth in a breaststroke-like motion. Accurate probing of larger *Artemia* was not feasible due to the presence of additional swimming appendages and an increased swimming strength in combination with a flexible curved anatomy that enabled them to overcome and escape MFS suction (**Supplementary Movie 3**). High-speed imaging experiments were also performed on free-swimming *Artemia* to allow for comparisons of the kinematics in tethered and free-swimming organisms. To analyse the secondary antennae motion which produces the *Artemia* propulsion, we have used the open-source, deep-neural network-based software DeepLabCut for markerless pose estimation (*61*) (**Fig. 3A**, **Supp. Fig. S3**, **Supplementary Movie 4–5** and *Methods*). The swimming frequency of *Artemia* decrease with $L$ (**Fig. 3B**), consistent with other studies (*29*, *49*). As *Artemia* develops, its antennae increase in length and width (**Supp. Fig. S4**), which translates into stronger propulsive limbs. By tracking the "armpit" and "elbow" angles $\theta_\mathrm{a}$ and $\theta_\mathrm{e}$ (**Fig. 3A** and **Supp. Fig. S5**) and their average amplitudes (**Fig. 3C–D**) we can describe their swimming kinematics. No difference in the kinematics is observed between the tethered and free-swimming experiments. By performing particle image velocimetry measurements, the fluid dynamics around the actively moving antennae are also found to be comparable in both scenarios, suggesting that the presence of the MFS does not influence the flow dynamics (see **Supp. Fig. S6** and *Methods*).

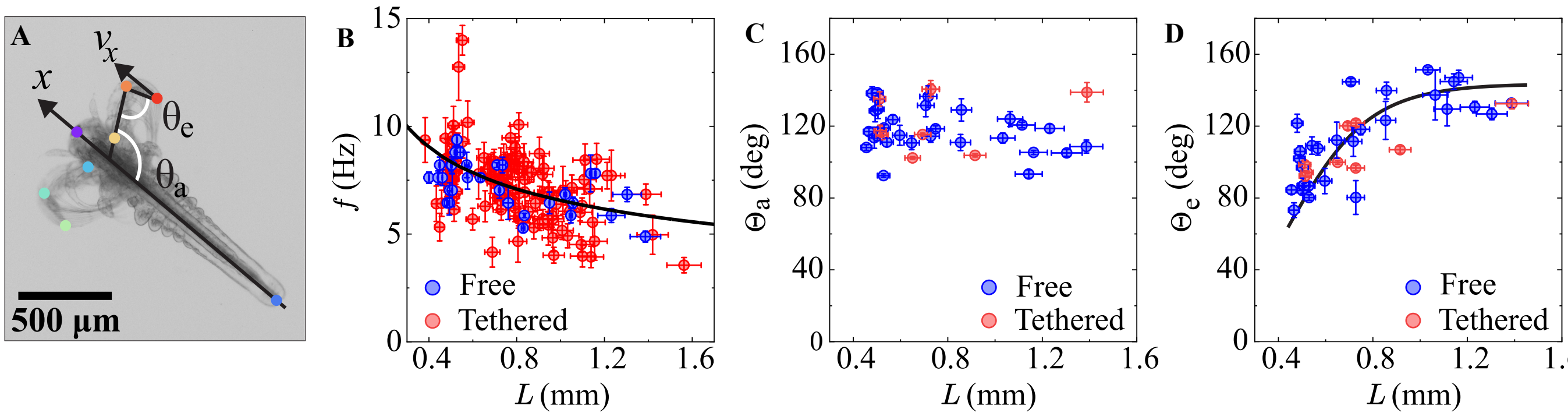

**Fig. 3 | Swimming kinematics. A)** *Artemia* swims by moving its antenna forward and backward (antenna tip velocity $v_x$ in the $x$-direction). Using DeepLabCut, we track the positions of 8 body parts (marked with coloured circles) as a function of time. The "armpit" and "elbow" angles $\theta_\mathrm{a}$ and $\theta_\mathrm{e}$ of the antennae are defined as shown in the image. **B)** The swimming frequency decreases as a function of body length. The solid line is the best fit to the data $f = (6.6 \pm 0.4) \cdot L^{-0.35 \pm 0.12}$ (with $L$ in units of mm). **C)** The average of the "armpit" angle amplitude remains constant as a function of body length, whereas **D)** the average of the "elbow" angle amplitude increases with body length. This increase (solid line to guide the eye) indicates that the antennae become more flexible, which allows for a more efficient movement where the recovery stroke can be performed with a bent limb that offers less resistance. The frequencies and antennae angle amplitudes are similar for free (blue data) and tethered (blue data) swimmers, which confirms that the MFS technique is minimally disruptive on the animal movement. The error bars in B–D are the standard errors from many swimming cycles of one individual.

***Force patterns and swimming dynamics***

The swimming force $F_{\mathrm{swim}}$ from an MFS experiment on a newly hatched *Artemia* is plotted as a function of time in **Fig. 4A**. The swimming cycle is a sequential repetitive motion of the secondary antennae that is comprised of a power and recovery stroke (**Fig. 4B**). To obtain a net forward displacement, *Artemia* must produce a positive mean propulsive force $F_{\mathrm{p}}$ (**Fig. 4A**). Other methods used in the literature to quantify swimming dynamics are the mean peak-to-peak force $F_{\mathrm{peak}}$ (**Fig. 4A**) as well as the mean integrated force (*5*). We have probed 129 different individuals of different lengths (three examples in **Fig. 4C**) and measured $F_{\mathrm{p}}$ (**Fig. 4D**) and $F_{\mathrm{peak}}$ (**Supp. Fig. S7A**) as a function of $L$. The measured $F_{\mathrm{p}}$ is independent of the MFS spring constant used (**Supp. Fig. S7B**). We find a linear correlation between the two forces: $F_{\mathrm{peak}} = (16 \pm 1) \cdot F_{\mathrm{p}}$ (**Fig. 4E**), which offers quantitative insights on the evolution of the relationship between power and recovery strokes in developing *Artemia* (represented by $F_{\mathrm{p}}$) while increasing the $F_{\mathrm{peak}}$. Both propulsive forces increase with body size as $F_{\mathrm{p}} \sim F_{\mathrm{peak}} \sim L^2$. The same scaling law has been reported for the nematode *C. elegans* (*43*) and for copepods (*5*). In the seminal work by Hill (*62*), the maximum force of a muscle scales as $F_{\mathrm{muscle}} \sim \sigma_0 L^2$, where $\sigma_0$ is the maximum isometric force per unit cross-sectional area (*29*), resulting in the observed $L^2$-scaling in the produced swimming force.

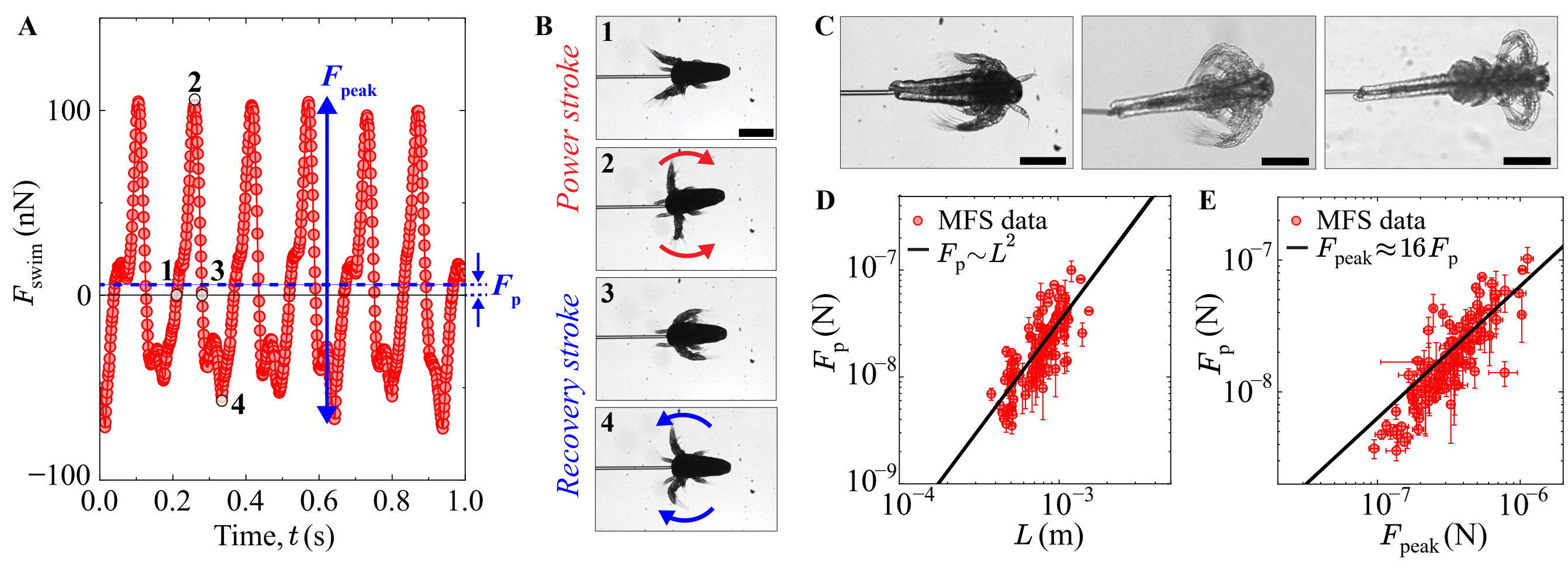

**Fig. 4 | Swimming dynamics. A)** Swimming force as a function of time of an early *Artemia* larvae ($L = 520 \pm 30\ \mu$m). Whereas the temporal variation of the pipette deflection is close to sinusoidal with a slight positive offset (**Fig. 2B**), the swimming forces shows a more complicated pattern. The mean propulsive force ($F_{\mathrm{p}}$) and peak-to-peak force ($F_{\mathrm{peak}}$) are calculated by averaging over many swimming cycles. **B)** At the beginning of the power stroke (1), the swimming force is zero and the antennae are extended frontally to the *Artemia* head. The antennae move outward, reaching a maximum force when the limbs are outstretched and perpendicular to the animal axis (2). The antennae move all the way back to the body and the swimming force reduces to zero (3). Here, the recovery stroke starts with the antennae moving frontally, rendering a negative swimming force. The force reaches a minimum value in (4), and towards the end of the recovery stroke, the antennae return to the initial position. **C)** Examples of three differently sized *Artemia* probed in this work. **D)** Mean propulsive force as a function of *Artemia* body length. The solid line is $F_{\mathrm{p}} = (0.031 \pm 0.003) \cdot L^2$, where the error is the 95% confidence interval of the fit. **E)** Mean propulsive force as a function of peak-to-peak force of *Artemia*. Despite rigorous standardization of the rearing and experimental conditions, unavoidable biological and procedural fluctuations introduced variability, manifesting as the observed dispersion in the data. The solid line is $F_{\mathrm{peak}} = (16 \pm 1) \cdot F_{\mathrm{p}}$, where the error is the 95% confidence interval of the fit. The error bars in D and E are the standard errors over many swimming cycles and the standard deviation of several measurements of the body length. Scale bars 300 μm. The scale bar in Panel B1 also applies to Panels B2–4.

The propulsive forces can also be described from a fluid mechanics perspective, where the hydrodynamic drag force acting on a body moving in fluid is given by $F_\mathrm{d} = \frac{1}{2} C_\mathrm{d} \rho S U^2$ where $C_\mathrm{d}$ is the drag coefficient (which depends on Re), $\rho$ the fluid density, $S$ the frontal surface area of the swimmer, and $U$ the characteristic swimming speed (*26*). For a living organism with complicated and time-varying morphology, it is not possible to accurately calculate the drag coefficient and frontal surface area. To find an order-of-magnitude correct scaling law for a living organism, we follow the recent theoretical work by Ventéjou *et al*. (*24*), predicting $F_\mathrm{d} \sim \eta f L^2$ in the Stokes regime. Since the swimming frequency of *Artemia* decreases only slightly with increasing $L$ (**Fig. 3B**), we find good agreement with $F_\mathrm{d} \sim L^2$ in **Fig. 4D**. To investigate this in greater quantitative detail, we plotted data only from the cleanest MFS experiments, where the organism was perfectly aligned with the tip of the cantilever and swimming in the exact focus plane of the microscope. Here, the scatter of the experimental data could be greatly decreased, and we find that $F_\mathrm{d} \sim \eta f L^2$ does not properly describe the *Artemia* swimming dynamics (**Suppl. Fig. S8**). A more careful description of the swimming kinematics and time-reversal symmetry breaking is needed for a correct model.

***Time reversal symmetry breaking***

Work with rudimentary artificial swimmers show that meso-swimmers can move in a reciprocal way, that is, without breaking time-reversal symmetry (*13*, *16–21*). In reciprocal motion, the path traced by the locomotory limb is identical to the path traced under time-reversal (*3*). For *Artemia*, the antennae trace an ∞-like loop during a single swimming cycle (**Fig. 5A–B**), indicating non-reciprocal motion. We define the area enclosed by this loop as the symmetry breaking area $A_\mathrm{sb}$ (**Fig. 5C**) and use it to quantify the non-reciprocity of the motion (*63*). For reciprocal motion, $A_\mathrm{sb} = 0$. For non-reciprocal motion, at least two degrees of freedom are required (*3*), which *Artemia* achieves by the elbow-like "joint" in its antenna. The symmetry breaking area of *Artemia* is non-zero and increases with $L$ (**Fig. 5D**). From our observations, two factors can affect $A_\mathrm{sb}$ of *Artemia.* In the power stroke, the moving antenna is passively bent back by the drag force of the opposing fluid. The bending force changes in a non-trivial manner along the length of the antenna due to the varying translational speed and cross-sectional area. The bending force will increase with the age (size) of the organism, whereas the bending stiffness of the antenna should increase strongly with antenna diameter ($EI \sim d_\mathrm{a}^4$ for a cylinder of diameter $d_\mathrm{a}$), making it more stable against bending. However, as *Artemia* develops, the antenna incorporates increasingly flexible articulations, leading to a reduction in effective bending rigidity. We observe that the antenna of the larger *Artemia* bend more under the opposing drag of the fluid, rendering an arc of a slightly smaller radius (than $l_\mathrm{a}$) in the contour drawn out by the antenna tip (**Fig. 5C**). This does not contribute significantly to the final symmetry breaking area. In the recovery stroke, the active bending of the antenna allows for a big difference in antenna tip motion as compared to the power stroke. This flexibility is enhanced and utilised more as the *Artemia* develops and drastically affects the symmetry breaking, especially in the second half of the power stroke and first half of the recovery stroke (left side of the ∞-shape in **Fig. 5A**). There thus seems to be a structural asymmetry in the anatomy of the antenna, allowing for more rigidity in the power stroke and active bending in the recovery stroke.

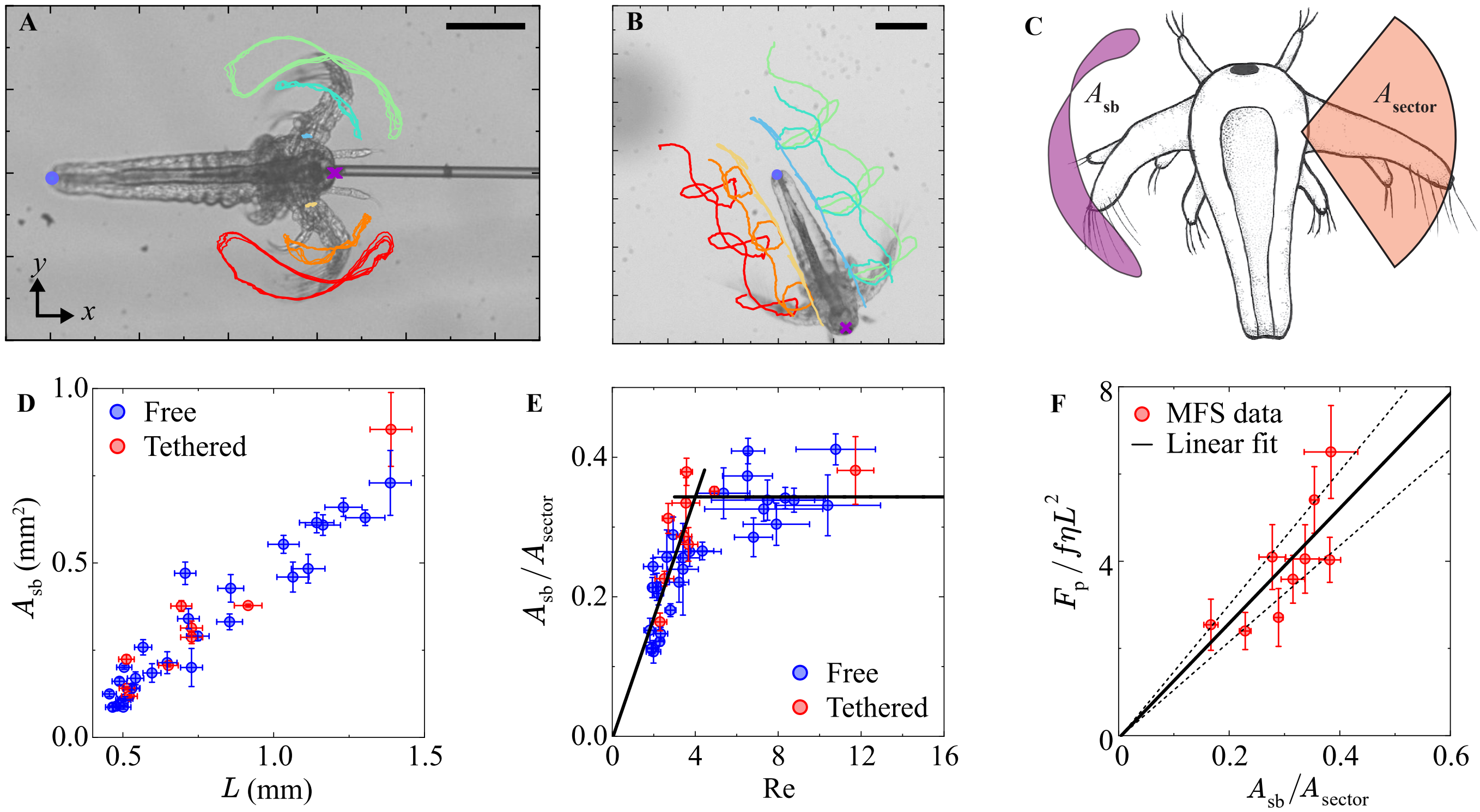

**Fig. 5 | Time-reversal symmetry breaking.** Tracking positions on an *Artemia* in a **A)** tethered and **B)** free-swimming experiment over several swimming cycles. In A the non-reciprocal antennae tip trajectories are clearly seen. **C)** The symmetry breaking area $A_{\mathrm{sb}}$ (purple region) and sector area $A_{\mathrm{sector}}$ (orange area). The latter would be traced out by a straight antenna moving with a reciprocal motion. **D)** The symmetry breaking area of *Artemia* as a function of body length in tethered and free-swimming experiments. **E)** The level of non-reciprocity ($A_{\mathrm{sb}}/A_{\mathrm{sector}}$) in the motion of *Artemia* as a function of Reynolds number (calculated using the antenna length and average antenna velocity). The free-swimming data (blue data) are consistent with the trend predicted with the MFS experiments (red data). **F)** The non-dimensionalised propulsive force increases linearly with the level of non-reciprocity in the swimming motion. The solid line is $F_{\mathrm{p}}/f\eta L^2 = (13 \pm 2) \cdot A_{\mathrm{sb}}/A_{\mathrm{sector}}$, where the error is the 95% confidence interval of the fit. The error bars in D–F are described in *Methods*. Scale bars in A–B 250 µm.

Since a larger *Artemia*, by definition, draws a larger $A_{\mathrm{sb}}$ due to their longer antennae, we decouple the effect of $l_{\mathrm{a}}$ by normalising $A_{\mathrm{sb}}$ with the area of the reciprocal motion sector $A_{\mathrm{sector}} = \pi l_{\mathrm{a}}^2 \Theta_{\mathrm{a}}/360°$ (**Fig. 5C**). The ratio $A_{\mathrm{sb}}/A_{\mathrm{sector}}$ provides a size-independent measure of the level of non-reciprocity of the motion. We find that $A_{\mathrm{sb}}/A_{\mathrm{sector}}$ increases linearly as a function of Re until a critical value of Re $\approx 4$ (**Fig. 5E**). After this, although not strictly required by fluid mechanics at the mesoscale where inertia should allow for reciprocal swimming, the growing *Artemia* continues to swim with a constant level of non-reciprocity until Re $\approx 12$. At even higher Re, the organism undergoes big changes by developing and using several additional appendages along the trunk (thoracopods) to propel itself forward with a metachronal gliding gait (**Supplementary Movie 3**). The transition to a constant level of symmetry breaking in the butterfly swimming motion of *Artemia* can be used for predicting the onset of the major gait change which is made possible by the mesoscale fluid mechanics. This mesoscale transition range occurs at similar Re values as reported for flapping plates and shell-less pteropod molluscs: Re $\gtrsim 5 - 20$ (*21–23*). Due to the transition to metachronal swimming, we cannot say if $A_{\mathrm{sb}}/A_{\mathrm{sector}}$ would decrease at increaing Re due to the increased relaxation of the Scallop theorem. Another organism would be needed to measure the level of non-reciprocity used by living meso-swimmers further into the intermediate Re regime.

We can now compare the directly measured propulsive forces with the level of non-reciprocity measured in the *Artemia* motion. To do so, we again normalise the measured mean propulsive force following the Stokes regime estimation ($F_{\mathrm{p}} \sim \eta f L^2$) (*24*) and plot it as a function of the level of non-reciprocity, finding $F_{\mathrm{p}}/\eta f L^2 \sim A_{\mathrm{sb}}/A_{\mathrm{sector}}$ (**Fig. 5F**). The propulsive force thus increases when the animal performs a more non-reciprocal motion. We find that the same logic holds for describing the speed of free-swimming *Artemia* (**Supp. Fig. S9**): $U = 2(A_{\mathrm{sb}}/A_{\mathrm{sector}})fL$. This expression takes inspiration from Gazzola *et al*. (*28*), without the need to precisely quantify a swimming amplitude, which is complicated given the asymmetric motion of the *Artemia* antenna. Anatomically, the results in **Fig. 5F** are mostly a consequence of an increase in the antenna flexibility (**Fig. 3D**) coupled with an apparent asymmetry in the structural rigidity of the antenna, allowing for rigidity against the fluid drag force in the power stroke yet active bending in the recovery stroke. At Re $> 4$, the anatomy seems to remain self-similar as the *Artemia* grows larger, rendering a constant $A_{\mathrm{sb}}/A_{\mathrm{sector}}$ and swimming forces that only increase due to changes in $L$ and $f$. From a fluid and swimming mechanics perspective, the general understanding has been that a living swimmer can enhance its propulsive force by either growing bigger or swimming faster. Our quantitative results show that the model micro- to meso-swimmer *Artemia* also can enhance its propulsive force by moving with a higher level of non-reciprocity. Understanding how the level of time-reversal symmetry breaking affects the propulsive force is important both from a fundamental physics point of view, as well as for designing better artificial swimmers, such as biomimicry meso-robots of the future.

***Universal swimming dynamics***

Finally, we conducted a broad literature review for directly measured swimming forces of a wide range of different micro- to meso-organisms (*5*, *10*, *38–43*, *47*). The resulting $F_{\mathrm{peak}}$ are plotted in **Fig. 6A** as a function of $\eta f L^2$. All data collapse on the same line, confirming the results based on swimming speed at the micro- to mesoscale regime (*24*). The scatter of the data in this universal scaling graph is due to several aspects. Firstly, there is a significant variation in swimming strategies and level of symmetry breaking displayed by the different organisms. For example, the bacterium *E. coli* pushes itself forward by rotating helical flagella, the alga *C. reinhardtii* uses two flagella to pull itself forward, the nematode *C. elegans* undulates its entire body, and *Artemia* and copepods use antennae to swim. Secondly, the swimmers have very different morphologies (e.g. cylindrical nematodes, round algae, and fusiform *Artemia*), all crudely simplified by one length scale $L$ in the model, which assumes that the length of the swimming appendages scales with $L$. Thirdly, different force measurement techniques and analysis strategies were employed and the forces reported for, e.g., bacteria are likely closer to a mean propulsive force and not $F_{\mathrm{peak}}$, whereas the copepods dynamics was reported as a mean integrated force. Nevertheless, the results of **Fig. 6A** confirms that meso-swimmers adhere to the same law as the micro-swimmers, even though Re $> 1$ and Stokes' law starts breaking down. In Svetlichny *et al*. (*5*), the copepods studied cover a range of Re $\approx 0.1 - 1000$ but still follow the low Re scaling model.

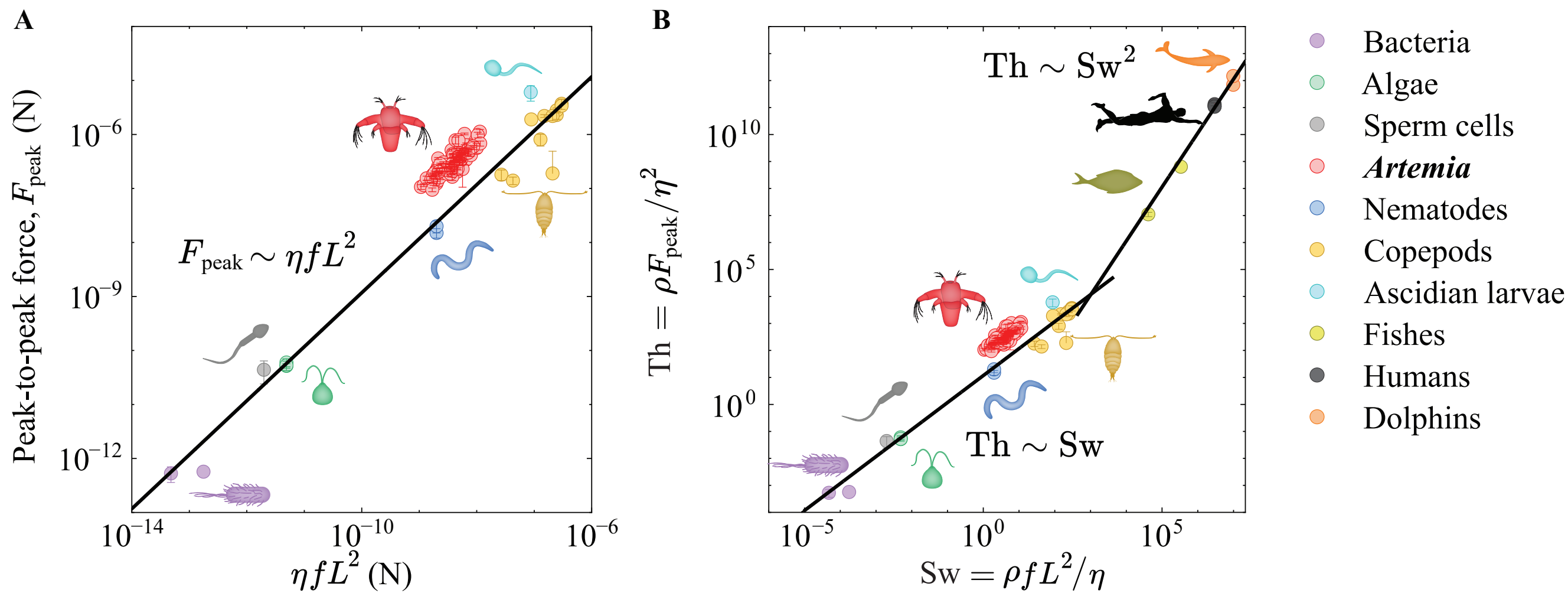

**Fig. 6 | Universal swimming dynamics. A)** The measured peak-to-peak force as a function the theoretical prediction for *Artemia* (this study), bacteria (*38*, *39*), algae (*40*, *41*, *47*), sperm cells (*42*), nematodes (*43*, *45*), *Ascidian* larvae (*10*), and copepods (*5*) swimming in water-like fluids ($\eta \approx 1$ mPas). The micro- to meso-swimmers follow the Stokes scaling of $F_{\text{peak}} = (12 \pm 2) \cdot \eta f L^2$ (solid line). The data for bacteria (*E. coli*) is a mean propulsive force, and the copepods data were in the cruising regime and reported as mean integrated forces, which are both similar in magnitude as $F_{\text{peak}}$. The algae (*C. reinhardtii*) data were reported as mean amplitudes, which we multiplied by 2 to get the $F_{\text{peak}}$. **B)** The thrust number as a function of swimming number for the data in A as well as for fishes (*64*, *65*), dolphins (*66*), and humans (*67*). The solid lines shows the expected scaling in the Stokes ($\text{Th} = (12 \pm 2) \cdot \text{Sw}$) and inertial ($\text{Th} = (0.011 \pm 0.002) \cdot \text{Sw}^2$) regimes (*24*). The errors for the fits are the 95% confidence intervals.

To better estimate when inertial forces start to influence the swimming dynamics, we include some direct and indirect force measurement results from the literature on fishes (*64*, *65*), dolphins (*66*), and humans (*67*) and plot the thrust number $\text{Th} = \rho F_{\text{peak}}/\eta^2$ (*24*) of all organisms as a function of swimming number $\text{Sw} = \rho f L^2/\eta$ (*28*) in **Fig. 6B**. We find excellent agreement between the data and the theoretical predictions of Ventéjou *et al*. (*24*), reporting a $\text{Th} \sim \text{Sw}$ scaling law in the Stokes regime, followed by a $\text{Th} \sim \text{Sw}^2$ scaling in the laminar and inertial regime (at ca. $\text{Re} > 10^2$). The force-based results show a similar transition to the laminar regime with organisms swimming at ca. $\text{Re} > 10^2 - 10^3$. More force-based data would be needed in the cross-over regime to probe the exact transition point. The propulsive force of a living organism is, within the scatter of the inherently noisy data of these biological systems, thus not sensitive to the addition of inertia at the mesoscale ($\text{Re} \sim 1 - 10^3$) but becomes apparent at $\text{Re} > 10^3$. This experimental finding partly goes against previous theoretical hypotheses of highly complex swimming dynamics at the mesoscale and confirms the recent results from swimming speed measurements on free-swimming organisms (*24*). The presented universal scaling laws can be conveniently used to roughly estimate the propulsive force of any swimming organism, regardless of size, morphology, or swimming motion. Future work with swimming meso-robots could provide cleaner data with insights into a potential cross-over regime between the Stokes and laminar scaling laws.

## Methods

### MFS manufacturing

Micropipette force sensors were made from hollow borosilicate glass capillaries (World Precision Instruments, USA, model TW100-6) with an outer/inner diameter of 1/0.75 mm, following the protocol by Backholm and Bäumchen (*46*). The capillary was heated, softened, and magnetically pulled apart using a Narishige PN-31 micropipette puller (Narishige International Limited, Japan). The settings for pulling were determined based on the desired characteristics and length of the pipettes. For the creation of a pipette with a length of 1.5–4 cm and a cantilever diameter of 10–30 µm, the chosen parameters were: main magnet 52, sub magnet 25, heater 90–95, and setscrew 5–8 mm. The end of the pulled micropipette was cut to a specific length using a MF2 microforge (Narishige International Limited, Japan) to obtain a smooth, open, and straight tip. This process involved the use of a heated filament with an affixed glass bead (filament heating was set to approximately 55–60 on the microforge). The bead was uniformly heated via the filament, and the pipette was brought into contact with it, this softened the glass and caused the pipette to adhere to the bead. Once adhesion occurred, the heating was immediately interrupted, and the sudden temperature gradient resulted in the pipette cleanly snapping just above the adhesion point between the pipette and the bead, forming a smooth tip rim. After cutting the pipette to the desired length, a right-angle bend was created at 900–1000 µm from the tip. The straight pipette was carefully placed onto a hot wire (with the filament heating set to 19–21 on the microforge), thereby thermally softening it. It was then gently nudged using a handheld metal wire until reaching a 90° angle, aided by the assistive grid of the microforge.

### Quasi-static MFS calibration

Micropipettes were calibrated following the water-droplet method (*46*, *55*). The pipette was mounted horizontally into the calibration setup and the bent section pointing vertically downwards (**Supp. Fig. S1A**). A syringe and tubing system were used to extrude a water droplet that rested on the outside of the pipette tip. The volume of the droplet was increased by injecting more water, causing the cantilever to deflect. This deflection was recorded at 24 fps using a camera (Canon EOS 90D with a macro lens, Canon Inc, Japan, and FLIR Grasshopper3 GS3-U3-23S6M-C, Teledyne Vision Solutions, USA), then analysed in MATLAB. By comparing the gravitational force of the ellipsoidal drop (density $\rho = 1000$ kg m$^{-3}$) of volume $V = \pi l_{\mathrm{min}}^2 l_{\mathrm{max}}/6$ (where the minor and major drop axis lengths $l_{\mathrm{min}}$ and $l_{\mathrm{max}}$ were determined through image analysis) with the micropipette deflection, the spring constant was determined as $k = \rho V g/x$ (**Supp. Fig. S1B**). Each cantilever was calibrated a minimum of 4 times, rendering an average value with a standard deviation.

### Dynamic MFS calibration

The dynamic calibration method was used to obtain the damping coefficient $b$ and the effective mass $m_{\mathrm{eff}}$ for the micropipettes. The experiments were conducted by deflecting the micropipette with a rigid L-shaped probe (**Supp. Fig. S2A**), with the entire system immersed in MilliQ water. The micropipette was imaged using a high-speed camera (AOS LPRI1000, AOS Technologies AG, Switzerland, and Phantom Miro 211 and T4040, AMETEK, Inc., USA) to record the deflection. Both the micropipette and the probe were mounted on a three-axial micromanipulator for precise positioning. A minimal contact point between the probe and micropipette tip ensured a clean release and was used to deflect the target pipette from its resting position, allowing the micropipette to oscillate and return to equilibrium. The damped harmonic oscillation model was fit to the time-deflection data (**Supp. Fig. S2B**) to obtain the parameters: undamped angular frequency $\omega_{\mathrm{o}}$, damping ratio $\xi$, initial amplitude $A$, and phase $\varphi$. The values for $b$ and $m_{\mathrm{eff}}$ were calculated as described in the *Results* section. Experiments were carried out for initial displacements ranging from 50 to 450 µm and conducted in brine and MilliQ mediums, both yielding consistent results (**Supp. Fig. S2C–D**). The calibration was done on 16 different cantilevers with lengths of $l_{\mathrm{MFS}} \approx 1.3 - 3.3$ cm and tip diameters of $d_{\mathrm{MFS}} \approx 26 - 42$ µm. In Frostad *et al.* (*58*), the effective mass is calculated by knowing the effective length, density, and the inner and outer diameter of their perfectly cylindrical cantilever. In our case, due to the tapered geometry of the micropipettes with both the inner and outer diameters decreasing along the length, and the effective length being unknown, it was difficult to calculate the effective mass analytically. With our dual calibration approach, however, we obtain the effective mass of the cantilever without needing to define the cantilever geometry.

**Experimental organisms and medium**

*Artemia* cysts were obtained from a commercial source (Reed Mariculture Inc. CA, USA) in a dehydrated state, preserved at 4 °C and used within the product's shelf-life. *Artemia* hatching and rearing adhered to the standardized protocols established by the Food and Agriculture Organization of the United Nations (FAO) and the International Artemia Aquaculture Consortium (IAAC) (*68*). Due to the absence of notochord, these organisms lack the neural complexity for a subjective sense of pain, and research on *Artemia* is currently exempt from the application of ethical authorizations.

For the assessment of motility, only individuals ranging from Instar I (newly hatched nauplius) to just before the differentiation stage, later Instar V, were used in the experiments. We selected this specific stage bracket because newly hatched *Artemia* nauplius primarily rely on the secondary antennae (**Supp. Fig. S10**) for propulsion. Despite featuring 3 pairs of specialized appendages, including primary (antennule) and secondary antennae and mandibles, which serve integrated roles in locomotion, sensory perception and feeding, the two other pairs apart from the secondary antennae, are comparatively smaller and do not exhibit significant motility that could contribute to net propulsion, thereby rendering the secondary antennae as the principal and sole propellers.

Artemia are protected by a thin, flexible chitinous exoskeleton composed of chitin, polysaccharides and proteins, forming an outer epicuticle and an inner fibrous procuticle, which provides structural integrity and protection while allowing for mobility. This exoskeleton is periodically shed through molting cycles, facilitating growth and morphological transitions. As the organism progresses through subsequent developmental stages, its body elongates and new appendages, called thoracopods, develop serially alongside the trunk (**Supp. Fig. S11**). These thoracopods, which differentiate from the thoracic segments, beat in a metachronal rhythm, eventually replacing the secondary antennae and becoming the primary motility appendages. Consequently, motility was assessed only during the stages where the secondary antennae function as the exclusive propulsive limbs in a breaststroke pattern.

These antennae are comprised of articulated ringlets that multiply and elongate through moulting during the larval development and are structurally organized into a main limb trunk, called exopod, and two branching ramifications, the endopod and the endite. These chitinous appendages form a flexible tubular structure. These structures are further equipped with long stiff quill-like structures, called setae, that aid in sensing, feeding and locomotion. Initially, the antennae contain a cellular medium, originating from the mesoderm, that progressively develops into muscle fibers, hemolymph, and connective tissue. This medium becomes densely innervated with neural fibers, enabling both sensory and contractile functions. Thus, locomotion is achieved through the coordinated contraction of antagonistic muscle groups, which act on the limb segments via a system of elastic tendons and springs. These muscles are anchored to the exoskeleton through tendinous cells and tonofibrillae, allowing for precise mechanical force transmission that ultimately enables the nauplii mobility and vital functions. However, the morphology of the antennae as well as the role of their structures on the swimming dynamics (**Supp. Fig. S12**) fall beyond the scope of this study.

The brine medium employed in the experiments, as well as in the hatching and rearing containers, were reconstituted from pure water (conductivity: 0.055 $\mu S\ cm^{-1}$, surface tension: 72.75 $mN\ m^{-1}$ at 20.5 °C). A commercially available blend of sea salts and minerals (InstantOcean, Spectrum Brands, USA) was added to this water in a ratio of 35–40 $g\ L^{-1}$ (35-40 ppt, specific gravity 1.024–1.028), maintaining a pH of 7.5–8.

**Hatching and sample collection process**

Saline medium was added to acrylic cylindrical hatcheries of 30 cm height and 6 cm diameter with a conical bottom (JBL GmbH & Co., Germany) in volumes of 250–300 mL per batch and coarse aeration provided from the bottom.

Columns were kept in a thermostatic room, half-submerged inside a glass tank (30x30x30 $cm^3$). Temperature was maintained between 25–28 °C inside the tank by using commercial aquarium thermostats. Hatcheries were illuminated by a full spectrum aquarium LED lamp at ≤2000 lux (JBL GmbH & Co., Germany). Light was provided in two different patterns: 1) uninterrupted illumination from 0–24 h to elicit hatching and 2) on a 12/12 h cycle for the remainder of the experimental phase.

Dehydrated cysts (Reed Mariculture Inc., CA, USA) were used at a concentration of 2 g $L^{-1}$. Cysts were directly added into the hatching columns for rehydration. To prevent aggregation, cysts were kept in continuous motion via coarse aeration. Incubation was conducted for a period of 24–26 h. After hatching, the aeration was interrupted, allowing the cysts to either sediment (unhatched) or float (empty shells). Newly hatched nauplii were separated from non-hatched and dead cysts by discarding the top and bottom phases of the hatching columns, then carefully collecting the rest into a beaker and transferred into the experimental arenas consisting of 10 cm petri dishes.

**Animal care and disposal**

After incubation, each batch was transferred into a sterile glass container and placed within the thermostatic main tank to allow them to grow and develop under a 12/12 h light cycle. Starting from the 48-hour mark, *Artemia* were fed a dehydrated phytoplankton blend comprised of *Dunaliella salina* (Fluxias GmbH, Germany), *Nannochloropsis* and *Tetraselmis* (Reed Mariculture Inc. CA, USA), both reconstituted in fresh saline medium. At the end of each experimental trial, test subjects were euthanized by immersion in 2% bleach saline solution for a minimum of 24 hours. After inactivation, the medium was responsibly disposed of in accordance with liquid waste management guidelines.

**MFS swimming force experiments**

We conducted our experiments using a pipette system that began with a 10 mL luer-lock plastic syringe to which an HI-7 injection holder (Narishige International Limited, Japan) was attached. The injection system consisted of a CI-1 connector leading to a 30 cm stretch of 1 mm Ø PTFE plastic tube, ending in a stainless-steel pipette holder. The pipettes were secured in the holder with a silicone gasket and an end cap.

Active swimmers were selected based on their positive phototaxis. This behaviour was induced by positioning a mini-LED pen torch (Lepro, Germany) on one side of a 10 cm petri dish, causing the *Artemia* to swim towards the light. Swimmers were carefully collected and transferred using Pasteur pipettes into the experimental arena. This process was carried out with minimal handling and allowing for a habituation period of at least 30 minutes before the beginning of the experimental trial.

Experimental arenas consisted of semi-closed enclosures crafted from optically transparent 2 cm deep and 10 cm diameter petri dishes (VWR International Oy, Finland). These dishes were covered with their own lids, which were sealed to prevent leaks. A 4x2 cm rectangular hole was carved into each lid, extending from the side and toward the center. This hole was used as the insertion point for the sample and to accommodate the micropipette cantilever, allowing for sufficient motion range for specimen capture and positioning. After the filtration and discard of the upper and lower phases, the experimental medium (typically 150–200 mL) was carefully transferred into a beaker and then to the experimental arenas, filling them completely and ensuring the medium contacted the lid. This was done to eliminate noise originating from the air-medium interface. Experimental animals were allowed to habituate to the arena for a minimum of 30 minutes. A full-spectrum LED light (GDWD Ltd., Latvia) was positioned to stimulate phototactic movement, creating a swarm on the right side of the arena, which served as the capturing area (**Supp. Fig S13**).

Meanwhile, the pipette holder and syringe system (Narishige International Limited, Japan) was set up to be actuated from on the left side of the experimental arena. The system was mounted on dampened posts (DP14A, Thorlabs, Sweden) which were attached to an active vibration isolation table (i4 medium, ST Instruments, Netherlands) and operated with micromanipulators (MN-153, Narishige International Limited, Japan). The micropipette holder was attached to the micromanipulator and the pipette placed so that the bent part was horizontal and aligned with the bottom of the petri dish. Using the micromanipulator, the micropipette was then carefully inserted through the rectangular hole at an angle of 10–20 degrees ensuring that the cantilever was fully submerged into the experimental medium to avoid disturbances from the meniscus and avoiding contact with the experimental arena.

The L-shaped pipette was brought close to the *Artemia* swarm, positioned just above the bottom of the petri dish. By applying gentle and continuous suction through the syringe, we captured individual *Artemia* as they swam along the bottom surface. Captured *Artemia* were confirmed to be unharmed and properly positioned with their body axis aligned with the cantilever, avoiding tilt and rotation (see examples of bad catches in **Supp. Fig. S14**). The captured individuals were then gently raised away from the bottom of the petri dish to prevent disturbances from unwanted contact and boundaries interactions with the Petri dish floor. During captures, we ensured that the direction of the force vector exerted by the swimmer was perpendicular to the cantilever. This meant capturing animals either at their telson (end of the abdomen) or at the top of their cephalothorax (between the primary antenna, at the naupliar eye position, **Supp. Fig. S10B**) with their axis aligned with the bent part of the cantilever.

Upon confirming a satisfactory capture, recording was conducted using a high- or semi-high-speed camera (FLIR Grasshopper3 GS3-U3-23S6M-C, Teledyne Vision Solutions, USA; AOS LPRI1000, AOS Technologies AG, Switzerland, and Phantom Miro 211, AMETEK, Inc., USA) at a framerate of 120–1000 frames per second. These recordings comprised 5-6 seconds of undisturbed swimming, enough to capture a minimum of 30 to 60 swimming cycles, followed by a gentle release of the individual by carefully pushing the syringe plunger and another 2-3 seconds of recording of the empty stationary pipette (**Fig. 2B**).

**Analysis of experimental swimming forces**

Using an in-house MATLAB code for image analysis the deflection $x$ of the MFS was captured. The swimming forces are calculated using the equation $F_{\text{swim}} = kx + b\dot{x}$ for MFS spring constant in the range $0.7 < k < 25\ \text{nN}\ \mu\text{m}^{-1}$ while neglecting the drag contribution if $k > 25\ \text{nN}\ \mu\text{m}^{-1}$. From the dynamic calibration data, the average drag coefficient ($b = 35.5\ \mu\text{Ns m}^{-1}$) is calculated for MFS with spring constants of $k < 25\ \text{nN}\ \mu\text{m}^{-1}$. The central differences for three deflection data points, that is the gradient, is used for calculating the velocity $\dot{x}$. These swimming forces are averaged over many swimming cycles to obtain the mean propulsive force $F_{\text{p}}$ (**Fig. 4A**). The peak-to-peak force is the difference between the maximum and minimum values of swimming force. This is averaged over many swimming cycle to obtain the $F_{\text{peak}}$. The tethered swimming frequency plotted in **Fig. 3B** was measured from the frequency of the pipette oscillation, whereas the free-swimming frequency was determined from the DeepLabCut analysis (see "Image analysis of *Artemia* kinematics"). The damping ratio and resonant frequency $f_{\text{r}} = \omega_o\sqrt{1 - 2\,\xi^2}/2\pi$ of the MFSs is plotted in **Supp. Fig. S2C–D**. No resonance occurs for $\xi \leq 1/\sqrt{2}$. If the resonant frequency matches with the swimming frequency of *Artemia* (~ 5–10 Hz) it can result in the amplification of the measured $F_{\text{swim}}$. Hence the MFS with resonant frequencies in this range ($5 < k < 8$ nN $\mu\text{m}^{-1}$) were not used for the swimming force experiments.

**Free swimming experiments**

The experiment was set up as described in *MFS swimming force experiments* but without the use of any pipettes. The motion of freely swimming *Artemia* was tracked using high-speed cameras (AOS LPRI1000, AOS Technologies AG, Switzerland, and Phantom Miro 211 and T4040, AMETEK, Inc., USA) at 700–1000 fps. Care was taken to only image swimmers that were at a sufficient distance from the bottom of the experimental arena to avoid drag or steric effects from the solid surface.

**Image analysis of *Artemia* kinematics**

The motion and structural change of the *Artemia*'s antenna is tracked using the DeepLabCut software. The software extracts a set of frames from the video (in .mp4 or .avi format) clip for the purpose of labelling. Depending on the quality of the initial tracking, a set of 60–120 frames are labelled manually which is used for creating the training dataset used by neural network ResNet-50 for tracking. Apart from the head and tail, three positions on the antenna (**Fig. 3A**, "shoulder", "elbow", and tip) are marked. We have used around 200–300 epochs and further increase in the value has not significant reduction in errors. The image size and number of frames used for analysis determine the time duration for completion of analysis. The output data provides the $x$ and $y$ coordinates of the tracked body parts. The *Artemia* is consistently moving and there is also a small change in the swimming direction. The sagittal line of the *Artemia* with respect to the $x$-axis is detected and then using the rotation matrix the marker positions are corrected for angle. The "head" or the "tail", depending on if it is a "head-catch" or a "tail-catch", of the *Artemia* is fixed to the origin. This method of correcting the angle and fixing the *Artemia* position is carried out for both tethered and free swimmers. The coordinates of the tracked body parts were transferred to the new coordinate system and then the time evolution of the armpit and elbow angles were calculated. Symmetry breaking area was also determined in the new frame of reference to eliminate effects of the rotation of the animal.

**Particle image velocimetry**

A stock solution of 3 μm polystyrene beads at a concentration of 50 mg ml$^{-1}$ (micromer® 01-00-303, micromod Partikeltechnologie GmbH, Germany) was used. An aliquot of 0.30 ml of this solution was suspended in 14.7 mL of the experimental medium arena, achieving a final concentration of 1 mg ml$^{-1}$. This setup provided a medium depth of ca. 4 mm, enabling imaging at our inverted microscope equipped with a 4x objective (IX73, Olympus, Japan). We employed a high-speed camera (Phantom T4040, Vision Research Inc., USA) capable of recording at a resolution of 2560x1664 pixels at 2000 frames per second. The spatial resolution was calibrated as 2.58 μm pix$^{-1}$.

**Requirements for MFS data to be included in Fig. 5**

In the DeebLabCut analysis on the tethered swimmers, we only include data from MFS experiments where the *Artemia* was swimming aligned to the plane of focus of the microscope to allow for accurate DeepLabCut analysis of the kinematics. Additionally, the swimmers used in **Fig. 6D–F** were perfectly aligned (defined as less than 10 degrees difference) to the end of the pipette to minimise the errors when correlating the swimming motion with the force measured with the cantilever in the $x$-direction. As a comparison, for the data in **Fig. 4D**, **4E**, and **6**, the organisms had an angle of less than 15 degrees with respect to the end of the pipette but were allowed to swim in a slightly different plane than the plane of focus as this does not affect the measured propulsive force.

**Error bars in Fig. 5**

The error bars for $A_{\mathrm{sb}}$ is the standard error of many swimming cycles probed in each experiment. The error for $L$ is the standard deviation measured several times on the same organism. The error bars for all axes in D and E are the propagated errors using the standard errors for $F_{\mathrm{p}}$, $f$, $l_{\mathrm{a}}$, antenna tip velocity, and $A_{\mathrm{sb}}$ from many swimming cycles, and the standard deviation of $L$. The error for $A_{\mathrm{sector}}$ is propagated using the standard error for $\Theta_{\mathrm{a}}$ over many cycles, and the standard deviation of $l_{\mathrm{a}}$ measured several times. The dashed lines in E are the 95% confidence intervals of the linear fit.

**Data Availability**

The datasets used for plotting all graphs in the paper and examples of raw data files are shared on Zenodo (https://doi.org/10.5281/zenodo.14513461).

**Code Availability**

The MATLAB analysis codes written for this paper are shared on Zenodo (https://doi.org/10.5281/zenodo.14513461).

**Acknowledgments**

This work was funded by the European Union (ERC, SWARM 101115076; M.B.) and the Research Council of Finland Fellowship (MesoSwim, 354904; M.B.). Views and opinions expressed are however those of the author(s) only and do not necessarily reflect those of the European Union or the European Research Council. Neither the European Union nor the granting authority can be held responsible for them. We thank Prof. Maciej Lisicki for valuable discussions.

**Author Contributions**

M.B designed the research and secured fund; R.A.L., S.N., and E.R. managed animal husbandries; R.A.L. and S.N. and performed the swimming force measurements; E.R. performed free-swimming experiments; S.N., A.H. and M.B. developed the dynamic calibration technique; S.N. and A.H. performed the dynamic calibration experiments; S.N., L.A. and E.R. performed the kinematics analysis; S.N. and G.M.B. performed the particle image velocimetry experiments; M.B., S.N., R.A.L., A.H., E.R. and L.A. analysed the data; and M.B., R.A.L. and S.N. wrote the manuscript.

**Competing Interests**

The authors declare no competing interests.